\documentclass[english,prb,aps,twocolumn,10pt]{revtex4-2}

\usepackage{amsmath}
\usepackage{amssymb}
\usepackage{wasysym}
\usepackage{graphicx}
\usepackage{babel}
\usepackage{array}
\usepackage{verbatim}
\usepackage[colorlinks=true, pdfstartview=FitV, linkcolor=blue, citecolor=blue, urlcolor=blue]{hyperref} 
\usepackage{xcolor}

\def\eg       {{\it e.g.}}
\def\ie       {{\it i.e.}}

\def\cf     {{\it cf.}}

\newcommand*{\tran}{^{\mathsf{T}}}

\renewcommand{\Re}{\operatorname{Re}}
\renewcommand{\Im}{\operatorname{Im}}

\newcommand{\ee}[1]{\cdot10^{#1}}
\newcommand{\mr}[1]{\mathrm{#1}}
\newcommand{\unit}[1]{\,\mathrm{#1}}
\newcommand{\um}{\,\mu{\rm m}}
\newcommand{\us}{\,\mu{\rm s}}
\newcommand{\uT}{\,\mu{\rm T}}

\newcommand{\ye}{\gamma_\mr{e}}

\newcommand{\var}[1]{\sigma^2_{#1}}

\newcommand{\fc}{f_\mr{c}}
\newcommand{\fo}{f_0}
\newcommand{\fmod}{f_\mr{mod}}

\newcommand{\ftf}{f_\mathrm{TF}}

\newcommand{\dfo}{\delta\!f_0}
\newcommand{\dfor}{\delta\!f_0^\mr{rand}}

\newcommand{\dfwin}{\Delta\!f_\mr{win}}
\newcommand{\dt}{\Delta t}

\newcommand{\rect}{\mathrm{rect}}
\newcommand{\Ro}{R_0}
\newcommand{\sinc}{\mathrm{sinc}}

\newcommand{\SR}{\mr{SR}}
\newcommand{\tint}{t_\mr{int}}
\newcommand{\tk}{t_k}

\newcommand{\dphi}{\delta\phi}
\newcommand{\eps}{\epsilon}

\begin{document}
	
	\title{Fast scanning nitrogen-vacancy magnetometry by spectrum demodulation}
	
	\author{P.~Welter$^{1,2,\dagger}$, B.~A.~J\'osteinsson$^{1,2,\dagger}$, S.~Josephy$^{2}$, A.~Wittmann$^{3}$, A.~Morales$^{2}$, G.~Puebla-Hellmann$^{2}$, and C.~L.~Degen$^{1,4}$}
	\affiliation{$^1$Department of Physics, ETH Z\"urich, Otto Stern Weg 1, 8093 Z\"urich, Switzerland.}
	\affiliation{$^2$QZabre AG, Regina-K\"agi-Strasse 11, 8050 Z\"urich, Switzerland}
	\affiliation{$^3$Institute of Physics, Johannes Gutenberg Universit\"at Mainz, Staudingerweg 7, 55128 Mainz, Germany.}
	\affiliation{$^4$Quantum Center, ETH Z\"urich, 8093 Z\"urich, Switzerland.}
	\email{degenc@ethz.ch}
	\thanks{$^\dagger$These authors contributed equally.}
	
	\begin{abstract}
	We demonstrate a spectrum demodulation technique for greatly speeding up the data acquisition rate in scanning nitrogen-vacancy center magnetometry.  Our method relies on a periodic excitation of the electron spin resonance by fast, wide-band frequency sweeps combined with a phase-locked detection of the photo-luminescence signal. The method can be extended by a frequency feedback to realize real-time tracking of the spin resonance. Fast scanning magnetometry is especially useful for samples where the signal dynamic range is large, of order millitesla, like for ferro- or ferrimagnets.  We demonstrate our method by mapping stray fields above the model antiferromagnet $\alpha$-Fe$_2$O$_3$ (hematite) at pixel rates of up to $100\unit{Hz}$ and an image resolution exceeding one megapixel.
	\end{abstract}
	
	\date{\today}
	
	\maketitle

	\section{Introduction}
	
	The scanning nitrogen-vacancy (NV) magnetometer is a next-generation scanning probe microscope able to quantitatively map surface magnetic stray fields with sub-50-nm spatial resolution~\cite{degen08apl,balasubramanian08,rondin12,maletinsky12}.  The technique relies on a single, optically-readable defect spin embedded in a sharp diamond tip that is scanned over the sample of interest.  Scanning NV magnetometry exploits the principles of quantum metrology to reach very high sensitivities, leading to new opportunities in the imaging of weakly magnetic systems.  In the recent past, scanning NV magnetometry has been used to map the stray field of magnetic vortices and domain walls in ferromagnets~\cite{rondin13,tetienne14,tetienne15,velez19}, antiferromagnets~\cite{appel19,wornle19,wornle21,hedrich21,finco21} and multiferroics~\cite{gross17,chauleau20,lorenzelli21}, skyrmions~\cite{dovzhenko18,gross18,jenkins19,velez22}, superconducting vortices~\cite{thiel16,pelliccione16,scheidegger22}, and two-dimensional ferromagnetism~\cite{thiel19,sun21,fabre21}.
	
	In the most commonly used detection scheme, the spin resonance frequency $\fo$ of the NV center is tracked using continuous-wave optically detected magnetic resonance (cw-ODMR) spectroscopy, and later converted to units of magnetic field using the spin's gyromagnetic ratio ($\ye = 2\pi\times 28\unit{GHz/T}$) \cite{schirhagl14}.  In this scheme, the microwave excitation frequency is scanned slowly across the spin resonance and the resulting absorption line-shape, detected using a photo-luminescence (PL) measurement, fitted to extract the resonance position.  Although this scheme works well for slow acquisition speeds, the non-linear least squares fitting of the spectrum is computationally expensive and ill-suited for real-time performance past a few Hz.  On the other hand, the high PL of modern NV tips \cite{wan18,hedrich20} should allow measurements of magnetic fields at rates of $100\unit{Hz}$ or faster while maintaining a high sensitivity below $10\unit{\uT}$.  The possibility of acquiring a scan in a matter of minutes rather than hours or days is enticing, and would further bolster the versatility of the technique.
	
	Several concepts for speeding up image acquisition have been presented in the past.  These include qualitative approaches that rely on PL quenching~\cite{rondin12} or fixed-frequency excitation~\cite{balasubramanian08}, semi-quantitative approaches using multi-frequency excitation~\cite{haberle13}, resonance tracking~\cite{schoenfeld11}, \textit{a posteriori} field reconstruction~\cite{wang20apl}, or combinations thereof~\cite{wang20apl}.  The highest reported scan rates for single spin magnetometry are around $40\unit{pixels/s}$ in imaging~\cite{wang20apl} and $100\unit{samples/s}$ in stationary benchmarks \cite{schoenfeld11}.  For ensemble NV sensing, real-time field tracking up to several hundred Hz has been reported~\cite{acosta10apl}.
	
	In this work, we present a signal demodulation method that easily scales to sample rates of $100\unit{Hz}$ and beyond, yet is directly quantitative without the need for post-processing, and that can tolerate sudden jumps of the magnetic field.  Our method is based on periodic excitation of the spin resonance by fast, wide-band frequency sweeps and spectral demodulation of the resulting PL signal.  Real-time feedback can optionally be included to increase the dynamic range. 	We demonstrate our technique by imaging the magnetic surface texture and domain structure of an antiferromagnetic thin film at pixel rates of up to $100\unit{Hz}$.

	\section{Traditional resonance detection}
	\begin{figure}
		\includegraphics[width=0.99\columnwidth]{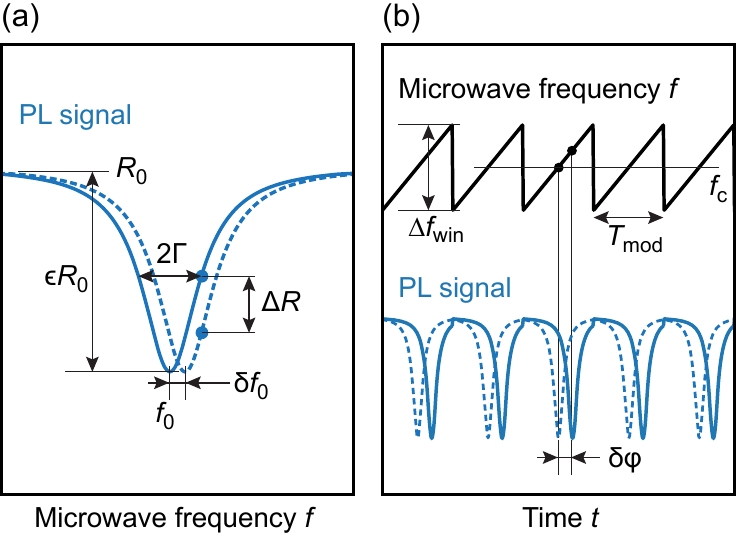}
		\caption{
			(a) In traditional ODMR spectroscopy, the resonance frequency $\fo$ is estimated by fitting the spectral peak.  Changes in magnetic field shift the position of the spectrum (dashed line).  $\Ro$ is the intensity of photo-luminescence (PL) emission, $\eps$ the spin contrast, and $\Gamma$ the linewidth parameter of the absorptive line shape.
			(b) In the spectrum demodulation technique, the spin resonance is excited periodically by fast microwave frequency sweeps (black saw-tooth curve).  The frequency modulation window $\dfwin$ is chosen larger than~$\Gamma$.  A change in the spin resonance by $\dfo$ then causes a phase shift $\dphi = 2\pi\dfo/\dfwin$ of the PL signal compared to the microwave drive (dashed line).  The phase shift is measured by demodulating the PL signal, \eg, by a lock-in amplifier.
		}
		\label{fig1}
	\end{figure}
	The canonical detection method in ODMR involves slowly sweeping the microwave excitation frequency across the spin resonance and monitoring the optical PL emission~\cite{gruber97}.  By fitting of the resonance curve with an appropriate line shape, most often Lorentzian,
	\begin{equation}
		R(f) = \Ro \left[1-\eps\left(1+\frac{[f-\fo]^2}{\Gamma^2}\right)^{-1}\right] \ ,
		\label{eq:lorentzian}
	\end{equation}
	the resonance frequency $\fo$ as well as other parameters including the resonance linewidth $\Gamma$, spin contrast $\eps$ and PL emission rate $\Ro$ can be extracted (Fig.~\ref{fig1}(a)).  To avoid the computationally expensive fitting, the change in resonance frequency $\fo$ can also be detected by observing the change in amplitude $\Delta\!R$ at a single frequency $f$ (blue dots in Fig.~\ref{fig1}(a)).  This `amplitude detection' can be extended to a few discrete frequency values to increase robustness \cite{kucsko13}.  Alternatively, a small (sub-linewidth) modulation of the microwave frequency or bias field can be applied to record a differential line shape or to frequency-lock to the resonance \cite{schoenfeld11, chechik16}.  The above procedures work well for analyzing spectra at a slow rate or for detecting and tracking small $\dfo<\Gamma$ changes in the resonance frequency.  However, they are ill-suited for real-time tracking of large spectral shifts.

	\section{Spectrum demodulation}
	
	In our spectrum demodulation method, the microwave drive $f(t)$ is swept quickly across a wide frequency window much larger than the resonance linewidth using a saw-tooth frequency modulation.  The modulation rate $\fmod$ is chosen much faster than the intended integration time per spectrum, yet much slower than the absorption and re-polarization rates of the spin.  The recorded PL signal as a function of time $R(t)$ is then a periodic concatenation of (truncated) resonance line shapes (Fig.~\ref{fig1}(b)).  By measuring the relative phase between the sawtooth drive $f(t)$ and the PL signal $R(t)$,
	\begin{equation}
		\phi = \frac{2\pi(\fo-\fc)}{\dfwin} \ ,
		\label{eq:phi}
	\end{equation}
	one can directly determine the frequency offset $\fo-\fc$.  Here, $\fc$ is the center frequency and $\pm \dfwin/2$ the frequency span of the sawtooth modulation.  Importantly, the phase $\phi$ is insensitive to the detailed line shape of the resonance.
	
	To experimentally determine $\phi$, we demodulate $R(t)$ at the modulation frequency $\fmod$ (for example, using a lock-in amplifier) and compute the argument of the in-phase and quadrature channels,
	\begin{align}
		a_1 &:= \langle R(t)\cdot e^{2\pi i \fmod t} \rangle  \ ,
		\label{eq:fundamental_demod} \\
		\phi &= \text{arg} \left(a_1\right) = \arctan(Y/X)  \ ,
		\label{eq:philockin}
	\end{align}
	where the angled brackets represent an average or a low-pass filter to reject the image at $2\fmod$.  $X=\Re(a_1)$ and $Y=\Im(a_1)$ are the in-phase and quadrature parts of the complex signal $a_1$, respectively.  The desired frequency $\fo$ then follows from Eq.~(\ref{eq:phi}),
	\begin{align}
		\fo = \fc + \frac{\phi\dfwin}{2\pi}  \ ,
		\label{eq:fo}
	\end{align}
	where the phase $\phi \in [-\pi,\pi[$.
	
	The demodulation can be extended to higher harmonics $n\fmod$ of $R(f)$, yielding a series of coefficients $a_n$.  A shift of the resonance frequency results in phase shifts of $n\phi$ for the harmonic of order $n$.  The expected amplitudes of higher harmonics are generally decreasing exponentially, therefore, phase measurements of the higher harmonics are increasingly noisy.  Nevertheless, the higher harmonics can be included in the analysis to obtain an improved estimate for $\fo$ (Appendices~\ref{appendix:sensitivity} and \ref{appendix:harmonics}).
	Furthermore, including harmonics up to second order allows extracting estimates for the resonance parameters $\Gamma$, $\epsilon$ and $\Ro$.  For a Lorentzian line shape, these are given by (see Appendix~\ref{appendix:sensitivity}):
	\begin{align}
		\Gamma &\approx \frac{\dfwin}{2\pi} \ln\left|\frac{a_1}{a_2}\right|  \ ,  \label{eq:gamma_approx}\\
		\eps &\approx \frac{\dfwin}{\pi\Gamma}e^{2\pi\Gamma/\dfwin}\left|\frac{a_1}{a_0}\right|  \ ,  \label{eq:eps_approx}\\
		\Ro &\approx a_0  \ . \label{eq:Ro_approx}
	\end{align}
	A more advanced analysis (Appendix~\ref{appendix:harmonics}) also exploits phase information from the higher order coefficients to improve the estimation of both the resonance frequency and the other parameters. However, in practice, Eqs.~(\ref{eq:gamma_approx}-\ref{eq:Ro_approx}) already provide decent estimates for the resonance parameters.
	
	\section{Implementation}
	\begin{figure}
		\includegraphics[width=0.99\columnwidth]{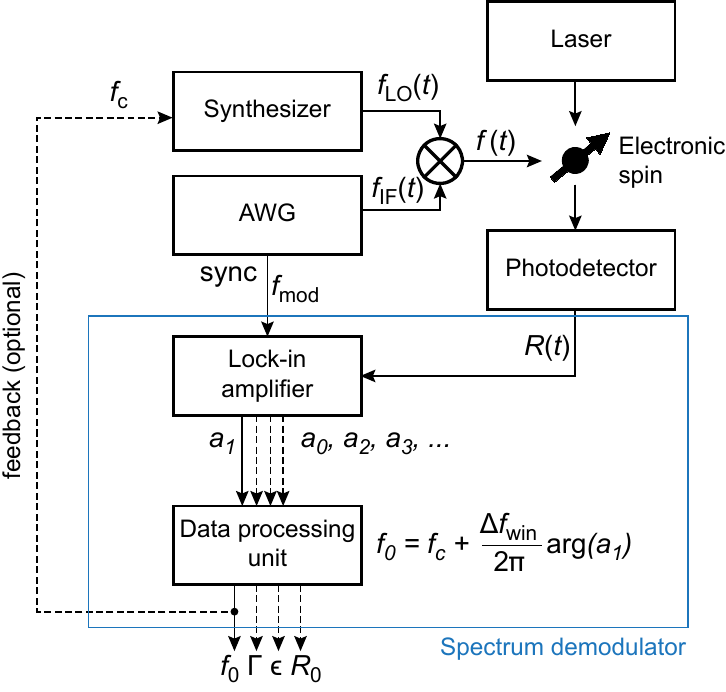}
		\caption{Block diagram of the spectrum demodulator.
			The complete system consists of conventional NV magnetometer setup to which a demodulator is added (dashed blue box).
			The demodulator can be implemented in a variety of ways and always includes a lock-in amplifier and data processing unit.  The lock-in amplifier computes the demodulated signals $a_n$, each representing the complex amplitude at the harmonic frequencies $n\fmod$.
			The data processing unit computes an estimate of the NV resonance frequency $\fo$ (and possibly other parameters including $\Ro$, $\epsilon$, $\Gamma$) from the signals $a_n$.
			Data processing can be a simple arithmetic operation [Eqs.~(\ref{eq:philockin},\ref{eq:fo})], or can use advanced methods such as Kalman or particle filters for improved sensitivity.
			Resonance tracking may optionally be implemented by feeding back the estimate of the resonance frequency $\fo$ to dynamically adjust the center frequency $\fc$ of the microwave modulation. In that case the spectrum demodulator together with the microwave synthesizer form a phase-locked loop (PLL) that locks onto the NV resonance frequency.
		}
		\label{fig2}
	\end{figure}

	A variety of ways may be devised to implement a spectrum demodulator.  The key elements of the system are shown in Fig.~\ref{fig2} and include: (i) the demodulator itself, (ii) a data processing unit for extracting the resonance parameters and (iii) optionally, a feedback to enable resonance tracking.
	In our system, the frequency modulation is generated on an arbitrary waveform generator, photo-detection achieved via a single-photon avalanche photo-diode (APD) and digital counter card, and all demodulation tasks are performed in software (see Section~\ref{sec:experimental}).
	
	Many other implementations can be considered: A fully analog system may combine a linear avalanche photo-diode (or a Geiger-mode APD with a down-stream low-pass filter) with a lock-in amplifier and a PID controller.  Conversely, a fully digital version may use a micro-processor to perform demodulation, signal extraction and tracking in a single unit.  Likewise, frequency modulation may be realized digitally (by direct digital synthesis) or fully analog (via a voltage-controlled oscillator).

	\section{Sensitivity}
	
	We next analyze the sensitivity of the spectrum demodulation technique and compare it to the conventional methods.  The main source of noise in the optical detection system is photon shot noise.  For low spin contrast $\eps \ll 1$, which is a good approximation for NV centers, the noise is Poissonian and white and the power spectral density is simply given by $S=\Ro$~\cite{rice15}.  Assuming a signal integration time of $\tint$, the equivalent noise bandwidth of the filter from Eq.~\eqref{eq:fundamental_demod} is $2/\tint$, and the variance of the demodulated signal is $\sigma^2=\Ro/\tint$.  This variance is evenly distributed over both quadratures, $\var{X}=\var{Y}=\Ro/2\tint$.
	
	Next, we use Eq.~(\ref{eq:philockin}) and Eq.~(\ref{eq:fo}) to convert the uncertainties in $X$ and $Y$ into an uncertainty $\dfo$ of the estimated frequency shift (see Appendix~\ref{appendix:sensitivity} for derivation).  The sensitivity $\eta$, defined as the uncertainty $\dfo$ normalized to unit time, is then given by:
	\begin{equation}
		\eta
		=\dfo\,\sqrt{\tint}
		\approx \frac{2\Gamma}{\eps\sqrt{R_0}} \times \frac{\alpha^2e^{\pi/\alpha}}{\sqrt{2}\pi^2} \ .
		\label{eq:sensitivity}
	\end{equation}
	Here, we introduce the relative window size $\alpha = \dfwin/(2\Gamma)$ as the ratio between $\dfwin$ and $2\Gamma$, and assume that $\alpha \gtrsim 1$.
	
	It is instructive to compare Eq.~\eqref{eq:sensitivity} to the optimum sensitivity figure for amplitude detection (Fig.~\ref{fig1}(a))~\cite{dreau11},
	\begin{align}
		\eta \approx \frac{2\Gamma}{\eps\sqrt{\Ro}} \times 0.77
		\label{eq:slope}
	\end{align}
	and to that of a least squares fit (Appendix~\ref{appendix:leastsquares}),
	\begin{align}
		\eta = \frac{2\Gamma}{\eps\sqrt{\Ro}} \times \sqrt{2\alpha/\pi}
		\label{eq:fit}
	\end{align}
	Clearly, for small $\alpha \rightarrow 1$, the sensitivity of all techniques is similar.  This is not surprising, because most frequency points lie in the vicinity of the resonance and contain useful information.  Conversely, for $\alpha \gg 1$, the sensitivity rapidly ($\eta \propto \alpha^2$) deteriorates for our spectrum demodulation technique because the signal power is increasingly distributed over higher harmonic coefficients $a_n$.
	In principle, the $\eta \propto \sqrt\alpha$ behavior of least-squares fitting can be recovered by including the $a_n$ in the analysis, however, this comes at the cost of increased analytical complexity (Appendix~\ref{appendix:harmonics}).  Overall, the window size $\alpha$ is an important parameter in our spectrum demodulation technique, because it provides us with a knob to balance between a large signal range (large $\alpha$) and a high sensitivity (small~$\eta$).
	
	\begin{figure}
		\centering
		\includegraphics[width=0.99\columnwidth]{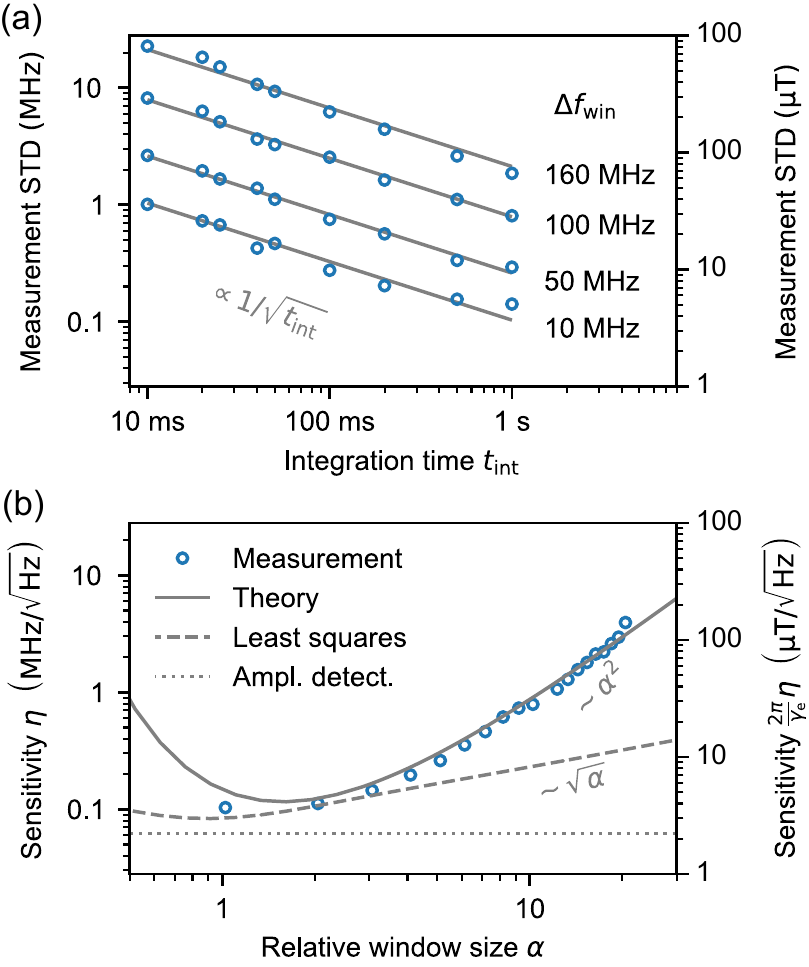}
		\caption{
			(a) Experimental standard deviation of the demodulator frequency output $\fo$ computed from 200 data points with no magnetic field modulation applied, plotted as a function integration time $\tint$ and window size $\dfwin$.  The solid lines act as guides to the eye, illustrating that the uncertainty scales as $\tint^{-1/2}$ [Eq.~\eqref{eq:sensitivity}].
			Right scale gives conversion to units of magnetic field. 
			(b) Experimental sensitivity vs.\ relative window size $\alpha = \dfwin/(2\Gamma)$.  The solid gray curve is the theory scaling [Eq.~\eqref{eq:sensitivity}].  The dashed gray curve is the theory scaling for least-squares fitting [Eq.~\eqref{eq:fit}].  The dotted gray curve is the theoretical limit for amplitude detection at the point of the steepest slope [Eq.~\eqref{eq:slope}]. 
			In this experiment, $\Ro\approx500\unit{kCt/s}$, $\eps\approx15\%$, and $2\Gamma\approx10\unit{MHz}$.
		}
		\label{fig3}
	\end{figure}

	\section{Tracking}
	
	To combine a high sensitivity with a large signal range, it is useful to include a tracking method that dynamically re-centers the frequency modulation window to the resonance position $\fo$, effectively forming a phase-locked loop (PLL).  At its simplest, we adjust the center frequency $\fc$ of the microwave drive to the last estimate for $\fo$ after every time step.  This method is not very robust, as a single noisy measurement can throw off the entire tracking, but we find it to be satisfactory in most of our experiments.
	Better approaches, not implemented here, would consider more than just the latest measurement. 
	One way is to implement a carefully tuned, higher-order PLL loop filter.  Another, digital approach is to make this filter itself adaptive, by use of recursive maximum-likelihood estimators that optimally consider previous time steps~\cite{moriyah10}.
	
	
	Ultimately, the choice of window size $\dfwin$ is a trade-off between signal-to-noise ratio (SNR) and tracking speed. The maximum frequency step allowed between two samples is given by $\dfwin/2$.  This implies a maximum tracking rate (slew rate) for frequency jumps of:
	\begin{align}
		\SR = \frac{\dfwin}{2\tint} \ .
		\label{eq:sr}
	\end{align}
	Using parameters typical for the experiments presented below ($\dfwin=30\unit{MHz}$, $\tint=10\unit{ms}$), the slew rate is approximately
	$\SR = 1.5\unit{GHz/s}$ corresponding to a magnetic slew rate of $\frac{2\pi}{\ye}\SR \sim 50\unit{mT/s}$.
	
	\begin{figure}
		\centering
		\includegraphics[width=0.99\columnwidth]{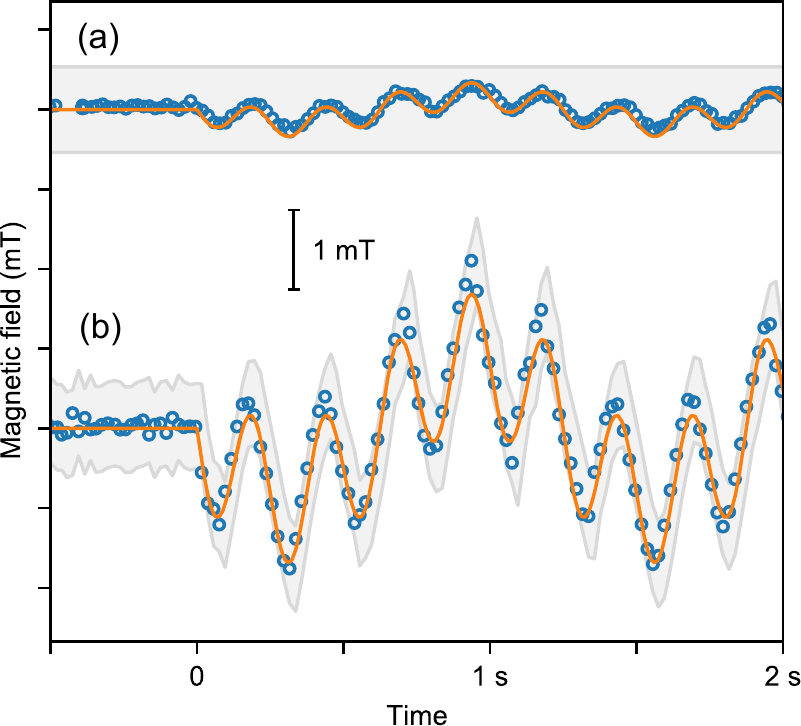}
		\caption{
			Real-time monitoring of magnetic field generated by passing a current through a coil underneath the NV probe.
			(a) With resonance tracking disabled.
			(b) With resonance tracking enabled.
			The waveform starts at $t=0\unit{s}$, and consists of two superimposed tones ($0.8\unit{Hz}$ and $4\unit{Hz}$) with equal amplitude. 
			Detection parameters are $\Ro\approx950\unit{kCt/s}$, $\epsilon\approx18\%$, $2\Gamma\approx12\unit{MHz}$, and $\dfwin=30\unit{MHz}$.
		}
		\label{fig4}
	\end{figure}

	\section{Experimental results}
	\label{sec:experimental}
	
	\begin{figure*}
		\centering
		\includegraphics[width=0.90\textwidth]{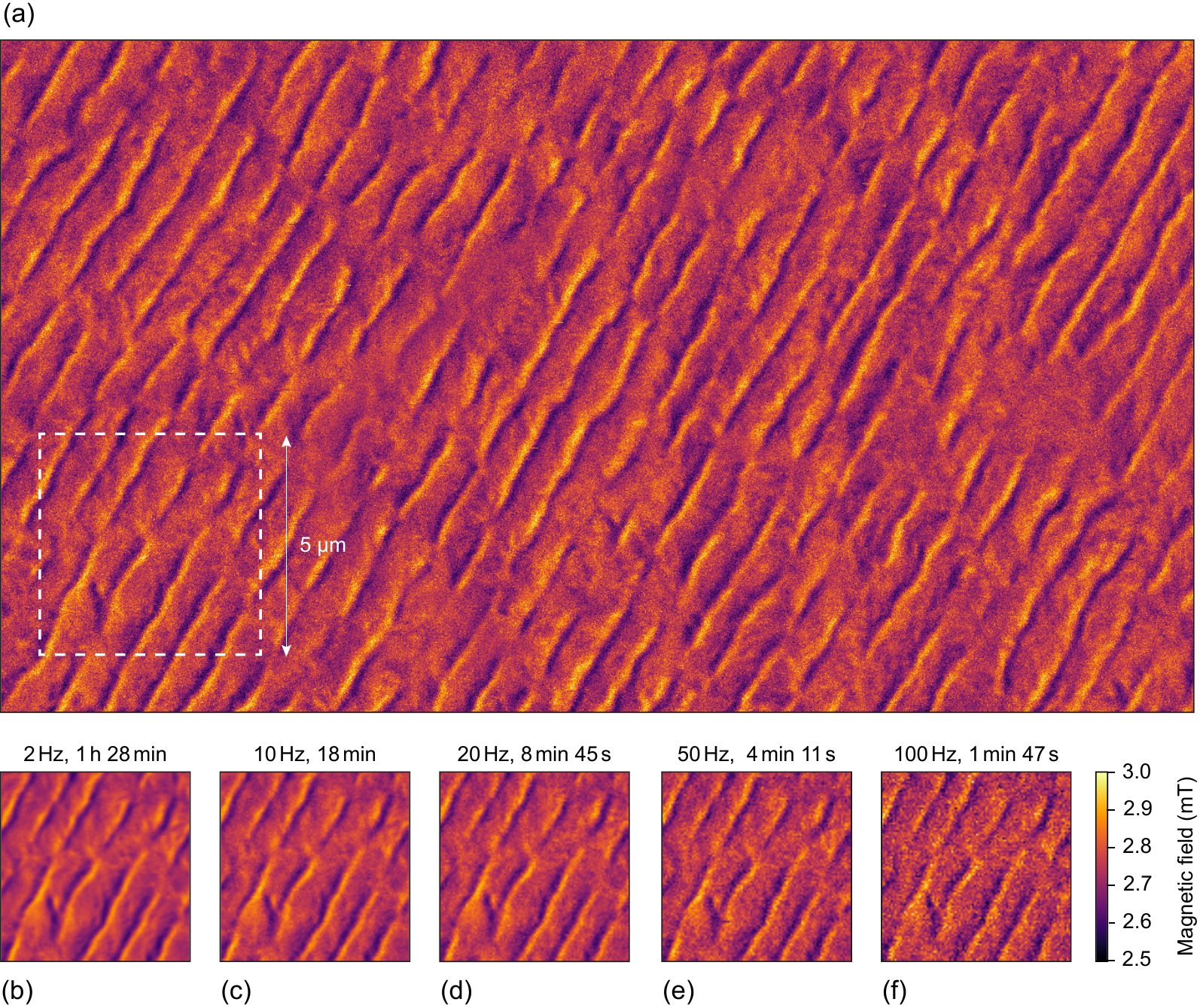}
		\caption{Magnetometry scans on $\alpha$-$\mathrm{Fe_2O_3}$ films using the spectrum demodulation technique.
			(a) Large-area scan ($27\unit{\um} \times 15.2\unit{\um}$) containing 1.03 million pixels. Pixel rate is $100\unit{Hz}$ and spacing is $20\unit{nm}$.  Total scan time is 3\,h\,5\,min.  The diagonal stripe pattern is likely due to steps in the surface topography~\cite{huxter22} while the large-scale structure reflects the magnetic domains.
			(b-f) $100\times100$ pixel scans recorded at varying rates, 2 to $100\unit{Hz}$, in the region corresponding to the dashed rectangle in panel (a). Titles indicate the scan rate and the total acquisition time.  Pixel spacing is $50\unit{nm}$.
			Experimental parameters are $\Ro\approx400\unit{KCt/s}$, $\eps\approx20\%$, $2\Gamma\approx11\unit{MHz}$, and $\dfwin=30\unit{MHz}$.
			A small bias field of ca. $2.75\unit{mT}$ is applied along the NV symmetry axis.
		}
		\label{fig5}
	\end{figure*}
	
	We experimentally demonstrate our spectrum demodulation technique using a commercial scanning magnetometer instrument (QSM, QZabre).  The scanning magnetometer is equipped with an arbitrary waveform generator and local oscillator (LO) for microwave control, and a single photon counting module for optical detection.
	We implement the frequency modulation by generating microwave chirp pulses at $\fmod = 1\unit{kHz}$ around a $100\unit{MHz}$ baseband frequency, and mix it with the LO to the desired $2-4\unit{GHz}$ final frequency centered at $\fc$.  We use an avalanche photo diode and a data acquisition card to count and bin the photons at a $20\unit{\us}$ dwell, corresponding to a data rate of $50,000$ points per second.  We demodulate the time trace in software by computing a fast Fourier transform of segments of duration $\tint$ and retaining the coefficients corresponding to $n\fmod$.  We then extract the resonance frequency $\fo$ from the phase of the first Fourier coefficient as per Eq.~\eqref{eq:fo}.
	Frequency tracking, when enabled, is implemented by updating the LO frequency according to the previously measured resonance frequency $\fo$.  In our current implementation, there is substantial latency associated with this process (up to $10\unit{ms}$), and in tracking mode we limit the maximum sample rate to $50\unit{Hz}$.  A future implementation will reduce this bottleneck by using an LO with lower latency.
	
	We start experiments by assessing the sensitivity of the method.  For this purpose, we record the resonance frequency $\fo$ at a constant rate given by $\tint$ for a total of 200 samples without applying a magnetic signal.  We then plot the standard deviation of the data record as a function of $\tint$.  Fig.~\ref{fig3}(a) shows the measured standard deviation for four window sizes $\dfwin$.  In all cases, the standard deviation scales with the inverse square root of $\tint$.  This is expected from Eq.~(\ref{eq:min_freq_fastscan}) and confirms that the measurement is limited by shot noise.  In Fig.~\ref{fig3}(b), we plot the sensitivity $\eta$ as a function of window size $\alpha=\dfwin/(2\Gamma)$.  The experimental data matches the sensitivity model from Eq.~\eqref{eq:sensitivity} exceptionally well.  Fig.~\ref{fig3}(b) also shows theory curves for least-squares fitting given by Eq.~\eqref{eq:fit} as well as the lower bound imposed by Eq.~\eqref{eq:slope}.
	
	To test the dynamic performance, we hover the scanning probe above a small copper coil ($\diameter 5\unit{mm})$ and feed a current waveform through the coil.  Fig.~\ref{fig4} presents traces of the resulting coil magnetic field recorded at a rate of $50\unit{Hz}$.  For reference, we also show the applied waveform on a matching scale (orange trace).  Fig.~\ref{fig4}(a) shows a trace recorded without tracking. Here, the dynamic range is limited to $\pm \frac12\ye\dfwin = \pm 0.53\unit{mT}$ by the chosen window size of $\dfwin = 30\unit{MHz}$.  Signals exceeding this range cannot be detected (not shown).  Fig.~\ref{fig4}(b) shows a corresponding trace with the tracking enabled.  The signal range is now much larger while the SNR is only marginally reduced (due to feedback latency). In both figures, gray bars indicates the instantaneous tracking window $\dfwin$.
	
	Whether or not the tracking should be enabled depends on the expected signal magnitude.
	If the expected signal dynamic range is small, less than approximately $\pm 30\unit{MHz}$ (equal to approximately $\pm 1\unit{mT}$, \ie{} $2\,\unit{mT}$ peak-to-peak), tracking is not necessary. The linewidth can be artificially broadened (or narrowed) by increasing (decreasing) the microwave power, such as to remain close to an optimum $\alpha \approx 3$.  By contrast, if the expected signal is strong ($\gtrsim \pm 30\unit{MHz}$), tracking is recommended. 
	
	
	Finally, Fig.~\ref{fig5} shows images of the stray field above a magnetic thin film obtained by scanning magnetometry. The sample is a $10\unit{nm}$ film of $\alpha$-Fe$_2$O$_3$ (hematite) grown epitaxially on an Al$_2$O$_3$ (001) substrate by off-axis magnetron sputtering capped with a $5\unit{nm}$ layer of Pt and a $2\unit{nm}$ layer of amorphous carbon. At room temperature, $\alpha$-Fe$_2$O$_3$ exhibits weak ferromagnetism due to the canting of the antiferromagnetically coupled magnetic sublattices within the easy plane. The average domain size is on the order of $1\unit{\um}$ making it a suitable materials system for our demonstration \cite{chmiel18}.
	%
	%
	The large image (Fig.~\ref{fig5}(a)) is $1350\times760$ pixels, recorded at rate of $100\unit{Hz}$ resulting in a total measurement time of 3\,h\,5\,min.  Despite the fast acquisition rate and the fairly weak signal ($\sim 500\unit{\uT}$ peak-to-peak), the image shows exceptional detail and a high SNR.
	Figs.~\ref{fig5}(b-f) show a $100\times100$ pixel sub-section of the image recorded at different rates 2 to $100\unit{Hz}$.  Although some loss in SNR becomes visible at high rates, features are well resolved in all images while the total acquisition time is dramatically reduced from 1\,h\,28\,min to below 2\,min.  For comparison, recording the large image at $2\unit{Hz}$ would have resulted in a scan time of approximately one week.

	\section{Outlook}
	In summary, we have introduced a technique for quantitative spin resonance frequency estimation that scales to at least 100 measurements per second.   Our method relies on a rapid, large-bandwidth frequency modulation of the microwave excitation and demodulation of the resulting PL signal.  Various trade-offs between accuracy (sensitivity) and speed (simplicity) are discussed.  We demonstrate fast scanning magnetometry by imaging the surface stray fields of a thin-film antiferromagnet at rates of up to 100\,Hz and sizes of up to one megapixel.
	
	Looking forward, the spectrum demodulation technique can be further improved in several directions.  A simple extrapolation (Appendix~\ref{appendix:maxrate}) indicates that the upper limit to the pixel rate is above $1\unit{kHz}$ for our experimental parameters, based on the available SNR.  In non-tracking mode, this speed is in principle accessible with our instrumentation.  In tracking mode, communication overheads currently limit the maximum rate to approximately $50\unit{Hz}$.  New hardware with reduced latency will likely improve this limit to well beyond $100\unit{Hz}$.  More advanced data processing and feedback techniques, such as the Kalman or particle filters, should further increase sensitivity and robustness and allow for even faster rates.  Also, data can be post-processed to optimize the SNR after a scan has completed.
	
	
	Another interesting future avenue is the implementation of gradiometry imaging.  Recent work has demonstrated spectacular improvements to sensitivity and image quality in scanning experiments by detecting the magnetic field gradient~\cite{huxter22}.  Gradiometry relies on a mechanical oscillation of the sensor above the sample surface, which up-converts the local gradient into a time-varying field set by the oscillation frequency.  While the original implementation relied on pulsed AC quantum sensing techniques, the concept can also be exploited in the context of the spectrum demodulation.  Assuming the mechanical oscillation frequency is $\ftf$, magnetic field gradients will lead to signal sidebands at $\ftf\pm n\fmod$ in addition to the original signal at $n\fmod$.  For typical tuning fork oscillators, $\ftf \sim 32\unit{kHz}$ is much larger than $\fmod = 1\unit{kHz}$, therefore, signals are spectrally well separated.  The gradient signal (and if desired, higher-order derivatives around multiples of $\ftf$) can be demodulated in exactly the same way as the standard demodulation technique (see Appendix~\ref{appendix:gradient} for details).  Measuring the gradient in addition to the direct field takes no extra measurement time, the only resource consumed is additional computation time. In the particular case where the scanning probe oscillates in the direction of the fast scanning axis, an estimate of the gradient (even if noisy) neatly integrates with a recursive estimator, improving the dynamic prediction of the field at the next pixel.

	
	%
	\begin{acknowledgments}
		The authors thank Marius Palm, Nils Prumbaum and Geoffrey Beach for discussions and support, and Larry Scipioni, Adam Shepard, Ty Newhouse-Illige, and James A Greer at PVD Products, Wilmington, Massachusetts 01887, USA for growing the $\alpha$-Fe$_2$O$_3$ thin film.
		This work was supported by the Swiss National Science Foundation (SNSF), Grant No. 200020\_175600, by the National Center of Competence in Research in Quantum Science and Technology (NCCR QSIT) of the SNSF, Grant No. 51NF40-185902, by Innosuisse Grant 43106.1 IP-ENG, and by the Advancing Science and TEchnology thRough dIamond Quantum Sensing (ASTERIQS) program, Grant No. 820394, of the European Commission.
	\end{acknowledgments}
	
%

	
	\appendix
	
	\section{Derivation of sensitivity}
	\label{appendix:sensitivity}
	
	We aim to compute the sensitivity to frequency shifts based on a single coefficient $a_1$. We start by deriving an approximate analytical expression for $a_1$.
	Assume zero frequency shift ($\fc = \fo$).  We define the sweep rate $v = \fmod\dfwin$ and the sweep period $T = 1/\fmod$. Over a single modulation period $|t|<T/2$, the luminescence time trace is simply the Lorentzian line shape
	\begin{equation}
		L(t) := R(v t + \fc) = \Ro (1-\eps[1+(v t)^2/\Gamma^2]^{-1}) \ ,
	\end{equation}
	as per Eq.~\eqref{eq:lorentzian}.  We can use this expression to describe the luminescence time trace over multiple periods,
	\begin{equation}
		R(t) = \left(\sum_{n=-\infty}^\infty \delta(t-nT)\right) * \Big(L(t)\,\rect(t/T)\Big) \ ,
		\label{eq:Rt_continuous}
	\end{equation}
	where $\rect(t/T)$ is the rectangular function which truncates the Lorentzian.  The convolution ($*$) with a delta comb then creates the periodic PL signal.
	
	We next compute the Fourier transform of Eq.~\eqref{eq:Rt_continuous},
	\begin{align}
		\mathcal{F}\left[R(t)\right](f)
		= & \left(\frac{1}{T}\sum_{k=-\infty}^{\infty}\delta\left(f-\frac{k}{T}\right)\right)  \nonumber \\
		& \cdot\Big(\mathcal{F}\left[L\right](f) * \left(T\,\sinc(Tf)\right)\Big) 
		\label{eq:FT_exact}
	\end{align}
	Here, we apply that multiplication and convolution are dual operations under a Fourier transform, that the Fourier transform of a comb is another comb, and that the Fourier transform of $\rect(t)$ is a $\sinc$ function, $\sinc(x) = \sin(\pi x)/(\pi x)$.
	The Fourier transform of the Lorentzian is a double-sided exponential, 
	\begin{equation}
		\mathcal{F}\left[L\right](f) = \Ro \delta(f) - \frac{\pi\Ro\eps \Gamma}{v} e^{-2\pi\Gamma |f|/v}  .
		\label{eq:ft_lorentzian}
	\end{equation}
	Note that we use the following convention for defining the Fourier transform:
	\begin{equation}
		\mathcal{F}[g](f) := \int_{-\infty}^\infty g(t) e^{-2\pi i ft}\mathrm{d}t.
	\end{equation}

	In Eq.~\eqref{eq:FT_exact}, for frequencies no larger than a small multiple of $\fmod$, the $\sinc$ function is much sharper than the decaying exponential in Eq.~\eqref{eq:ft_lorentzian}, and we approximate
	\begin{equation}
		T\sinc(Tf) \approx \delta(f) \ .
		\label{eq:truncation}
	\end{equation}
	%
	This approximation is equivalent to neglecting the truncation, noting that (i) the Lorentzian has largely decayed towards the side of the window, and (ii) the truncation creates a discontinuity that mostly contains higher frequencies ($\gg \fmod$) that are rejected by the demodulation.

	Combining Eqs.~(\ref{eq:FT_exact}), (\ref{eq:ft_lorentzian}) and (\ref{eq:truncation}), we find
	\begin{align}
		\mathcal{F}\left[R(t)\right](f) 
		= & \left(\sum_{k=-\infty}^{\infty}\delta\left(f-\frac{k}{T}\right)\right)  \nonumber \\
		& \cdot\Big(\frac{\Ro}{T} \delta(f) - \frac{\pi\Ro\eps \Gamma}{\dfwin} e^{-2\pi\Gamma |f|/v} \Big),
		\label{eq:FT_approx}
	\end{align}
	Here, the multiplication with the comb discretizes the spectrum of $R(t)$, as expected from its periodicity. Indeed, we can represent $\mathcal{F}\left[R(t)\right](f)$ as a Fourier series:
	\begin{equation}
		\mathcal{F}\left[R(t)\right](f) \overset{!}{=} \sum_{k=-\infty}^\infty a_k\delta(f-k/T).
	\end{equation}
	By comparison of coefficients, we find
	\begin{align}
		a_0 &= R_0 -\frac{\pi R_0\eps\Gamma}{\dfwin}\approx R_0    \label{eq:a0_approx}\\
		a_1 &\approx \frac{\pi\Ro\eps \Gamma}{\dfwin} e^{-2\pi\frac{\Gamma \fmod}{v}} = \frac{\pi\Ro\eps \Gamma}{\dfwin}e^{-2\pi\Gamma/\dfwin}    \label{eq:a1_approx}\\
		a_2 &\approx \frac{\pi\Ro\eps \Gamma}{\dfwin}e^{-4\pi\Gamma/\dfwin}    \label{eq:a2_approx}.
	\end{align}
	As expected, $a_1$ is purely real in the absence of a shift of $\fo$. Otherwise, by Eq.~\eqref{eq:phi}, $a_1$ is rotated in the complex plane by an angle $\phi$. Since we know the magnitude $|a_1|$ from Eq.~\eqref{eq:a1_approx}, as well as its variance from shot noise ($\var{X}=\var{Y}=\Ro/2\tint$, see main text), it is straightforward to compute the uncertainty in its phase, using standard error propagation,
	\begin{equation}
		\dphi
		=\frac{\sqrt{R_0/2\tint}}{|a_1|}  \ .
	\end{equation}
	Equivalently, the uncertainty in estimated resonance frequency (\cf{} Eq.~\eqref{eq:fo}) is given by
	\begin{equation}
		\dfo = \dphi \frac{\dfwin}{2\pi} = \frac{2\Gamma}{\eps\sqrt{R_0}}\frac{\alpha^2e^{\pi/\alpha}}{\sqrt{2}\pi^2}\left(\tint\right)^{-1/2} \ .
		\label{eq:min_freq_fastscan}
	\end{equation}
	In the last step we have introduced the relative window size $\alpha = \dfwin/(2\Gamma)$.  The sensitivity $\eta$, defined as the uncertainty in $\fo$ normalized to unit time, is given by
	\begin{equation}
		\eta = \dfo \cdot \sqrt{\tint}\ ,
	\end{equation}
	immediately yielding Eq.~\eqref{eq:sensitivity}.
	
	Solving Eqs.~\eqref{eq:a0_approx} to \eqref{eq:a2_approx} for $\eps$, $\Gamma$ and $\Ro$ yields Eqs.~\eqref{eq:gamma_approx} to \eqref{eq:Ro_approx}:
	
	\begin{equation}
		a_0 \approx R_0,
	\end{equation}
	\begin{align}
		\vert a_1/a_2\vert &= \exp\left(2\pi\Gamma/\dfwin\right) \nonumber\\
		\Rightarrow\Gamma &= \frac{\dfwin}{2\pi}\ln\left| a_1/a_2\right|,
	\end{align}
	\begin{align}
		\vert a_1/a_0\vert &\approx \frac{\pi\eps \Gamma}{\dfwin} e^{-2\pi\Gamma/\dfwin} =  \frac{\pi\eps}{2\alpha} e^{-\pi/\alpha}\nonumber \\
		\Rightarrow \eps &= \frac{2\alpha}{\pi}e^{\pi/\alpha}\left|\frac{a_1}{a_0}\right|.
	\end{align}
	
	\section{Higher-order coefficients}
	\label{appendix:harmonics}
	
	The phase of $a_1$ is a straightforward way to estimate the resonance frequency. But we can extend our analysis to include higher harmonics ($n\fmod$), allowing us to potentially improve sensitivity and extract information on $\Gamma$, $\eps$ and $\Ro$ at the same time.  In the following, we will move from a continuous-time picture to discrete time, and assume we have sampled $R(t)$ on a regular grid $t_k$, with a sampling time $\dt=\tint/N$ and number of samples $N$. The harmonic coefficients up to $n=2$ are given by:
	\begin{align}
		a_0 &= \frac1N \sum_k R(\tk) \\
		a_1 &= \frac1N \sum_k R(\tk) e^{2\pi i\fmod\tk} \label{eq:a1_dft}\\
		a_2 &= \frac1N \sum_k R(\tk) e^{2\pi i2\fmod\tk} \\
		\dots \nonumber
	\end{align}
	Note how we also include the DC value $a_0$. Eq.~\eqref{eq:a1_dft} is the discrete-time analogue of Eq.~\eqref{eq:fundamental_demod}.
	
	A shift of the resonance frequency that results in a phase shift of $a_1$ by $\phi$ will shift $a_2$ by $2\phi$, and so on. The expected amplitudes of higher harmonics are generally decreasing exponentially (Appendix~\ref{appendix:sensitivity}), so measurements of the phase of the higher harmonics are increasingly noisy. Indeed, all coefficients have the same variance $\sigma^2 = R_0/\tint$, and, by the central limit theorem, their distribution is closely normal.
	
	All of the coefficients $a_j$ carry usable information about all of the parameters. For instance, Eqs.~\eqref{eq:gamma_approx}--\eqref{eq:Ro_approx} clearly discard the phase information in $a_2$.
	The challenge is to find a suitable way to combine all of these measurements (real and imaginary part of each $a_j$) into a single estimate of all the parameters. Because all $a_j$ satisfy the condition of being normally distributed with equal variances, the maximum-likelihood estimate is in fact the one produced by least-squares optimization. That is, the optimum estimate is:
	\begin{equation}
		\hat f_0, \hat \epsilon, \hat \Gamma, \hat R_0 = \mathrm{argmin} \sum_j \lvert \tilde a_{j}- a_j \rvert^2,
		\label{eq:lstsq_demod}
	\end{equation}
	where $\tilde a_{j} = \tilde a_j(f_0, \epsilon, \Gamma, \Ro)$ represent the expected coefficients based on a model of the line shape, \cf{} Eqs.~\eqref{eq:a0_approx}--\eqref{eq:a2_approx}.
	
	With the only difference that the raw data are first Fourier transformed, Eq.~\eqref{eq:lstsq_demod} is equivalent to directly fitting the Lorentzian spectrum, a method we earlier dismissed as too computationally expensive. The key advantage here is that we can truncate the sum in Eq.~\eqref{eq:lstsq_demod} at perhaps $n=3$, as the remaining $a_j$ are small and the information they carry is minimal.  But indeed there is a trade-off to be made here, between computational complexity and sensitivity.

	\section{Derivation of sensitivity of least-squares fitting}
	\label{appendix:leastsquares}
	
	Next, we aim to compare the uncertainty above in Eq.~\eqref{eq:min_freq_fastscan} to the standard method of least-squares fitting the Lorentzian spectrum directly. A complete calculation of the error propagation through a full nonlinear least-squares fitting operation is intractable and unlikely to yield any tangible insights. Instead, to simplify the calculations, we relax the requirements by making the following assumptions:
	\begin{itemize}
		\item We only fit the resonance frequency $\fo$, and assume that the linewidth $\Gamma$, contrast $\eps$ and count rate $R_0$ are known exactly.
		\item We further assume that we already have a decent estimate of $\fo$, and we perform a single \emph{linear} least-squares step starting from the true value $\fo$.
		\item Likewise, the window shall be well-centered on the resonance, \ie{} $f_c\approx\fo$.
	\end{itemize}
	The model function which we fit to our data hence is
	\begin{equation}
		R_{\fo}(f) = \Ro (1-\eps[1+(f-\fo)^2/\Gamma^2]^{-1}).
		\label{eq:lorentzian_lstsq}
	\end{equation}
	%
	%
	Our measurement is a noisy sample of this function:
	\begin{equation}
		y_k = R_{\fo}(f_k) \frac{\tint}{N} + w_k, \quad k = 1, \ldots, N.
	\end{equation}
	Here, we sample the spectrum at the frequencies $f_k$. The total integration time $\tint$ is spread across all $N$ points. The random variable $w_k\sim\mathcal{N}(0, \sigma^2=\Ro\tint/N)$ captures the shot noise.
	
	The Jacobian of the least-squares problem reads:
	\begin{equation}
		J_k =  \frac{\partial R_{\fo}(f_k)}{\partial \fo}\frac{\tint}{N}.
	\end{equation}
	Note that because we are only fitting a single parameter $\fo$, the Jacobian is simply a row vector. Let $\hat\fo$ be the least-squares estimate of the resonance frequency. Its variance is given by:
	\begin{align}
		\mathbb{V}\left[\hat\fo\right]
		&= \sigma^2\left(J\tran J\right)^{-1}  \\
		&= \sigma^2 \left(\sum_{k=1}^NJ_k^2\right)^{-1} \\
		&= \sigma^2 \left(\sum_{k=1}^N\left(\frac{\partial R_{\fo}(f_k)}{\partial \fo}\,\frac{\tint}{N}\right)^2\right)^{-1}\\
		&\approx \sigma^2 \left(\int_{\fc-\alpha\Gamma}^{fc+\alpha\Gamma}\left(\frac{\partial R_{\fo}(f)}{\partial \fo}\,\frac{\tint}{N}\right)^2\frac{N\mathrm{d}f}{\dfwin}\right)^{-1}. \label{eq:variance_lstsq_int}
	\end{align}
	The integral runs over the frequency window. The approximation in the last step is valid when the frequency points are many ($N\gg1$) and equally spaced.
	
	Inserting Eq.~\eqref{eq:lorentzian_lstsq} into~\eqref{eq:variance_lstsq_int}, the integral can be solved analytically, finally yielding
	\begin{equation}
		\dfo=\sqrt{\mathbb{V}\left[\hat\fo\right]} =\frac{2\Gamma}{\eps\sqrt{\Ro\tint}}\cdot \sqrt{\frac{\alpha}{\frac{\alpha^4 +\tfrac{8}{3}\cdot\alpha^3-\alpha}{\left(1+\alpha^2\right)^3}+\arctan\alpha}}.
		\label{eq:lstst_uncertainty}
	\end{equation}
	where $\alpha$ is the relative window size.  This expression is minimized for $\alpha\approx1$.  In that case, the square root evaluates to approximately unity as well. The ultimate sensitivity with least-squares fitting is thus:
	\begin{equation}
		\eta_\mathrm{lstsq}^\mathrm{opt} = \dfo\cdot\sqrt{\tint}\approx \frac{2\Gamma}{\eps\sqrt{\Ro}},
	\end{equation}
	similar to the optimum single-point sensitivity of amplitude detection, see Eq.~\eqref{eq:slope} in the main text. Note that in practice, this sensitivity is out of reach by a small factor. The reason is that resonance frequency is \emph{not} the only quantity that must be estimated from the data. Robust estimation of also the contrast and the linewidth requires that the window is large enough to also capture the tails of the resonance ($\alpha\gtrsim2$).
	
	If the window is much larger than the linewidth ($\alpha\gg1$, as required by wide-dynamic-range measurements), the square root in Eq.~\eqref{eq:lstst_uncertainty} is $\sim\sqrt{2\alpha/\pi}$, meaning that the sensitivity is given by
	\begin{equation}
		\eta_\mathrm{lstsq} \approx \frac{2\Gamma}{\eps\sqrt{\Ro}} \cdot\sqrt{2\alpha/\pi}.
	\end{equation}
	This expression scales like $\sqrt{\alpha}$, which is intuitively plausible as only a fraction of the total integration time, of order $\tint/\alpha$, is spent sampling the spectrum near the actual resonance line.

	\section{Maximum rate}
	\label{appendix:maxrate}
	
	We estimate an upper bound for the maximum tracking rate solely limited by the SNR.  To get such a bound, we determine the integration $\tint$ where the uncertainty in $\dfo$ (according to Eq.~\ref{eq:min_freq_fastscan}) becomes equal to the estimate $\dfor$ one would obtain if the phase were completely random over $[-\pi,\pi[$.  Defining $\dfor$ by the square root of its variance,
	\begin{align}
		\dfor
		= \frac{\dfwin}{2\pi}\,\left[ \frac{1}{2\pi} \int_{-\pi}^{\pi} \mr{d}\phi\, \phi^2 \right]^{1/2} 
		= \frac{\dfwin}{\sqrt{3}} \ ,
	\end{align}
	setting $\dfo = \dfor$, and solving for $\tint^{-1}$ we find
	\begin{align}
		\tint^{-1} = \frac{\epsilon^2\Ro\pi^4}{6\alpha^2\exp\left(\frac{2\pi}{\alpha}\right)} \ .
	\end{align}
	Assuming numbers typical for our experiments ($\epsilon = 0.2$, $\Ro = 5\ee{5}\unit{s^{-1}}$, $\dfwin=30\unit{MHz}$, $\Gamma=5\unit{MHz}$), we find $\tint^{-1} = 4.4\unit{kHz}$.
	
	On the other hand, the NV center has a response time of $\sim 1\us$ \cite{schoenfeld11}. For accurate sensing, the sweep period must be much larger than this, \eg{} $\tint\gtrsim 100\us$. This equally imposes a limit on the maximum sampling of the same order of magnitude.

	\section{Demodulation of magnetic field gradient}
	\label{appendix:gradient}
	
	In the following, we show how demodulating the luminescence signal at $\ftf\pm n \fmod$ gives access to the magnetic field gradient along the cantilever oscillation axis.
	In the presence of a magnetic gradient, the probe experiences an additional AC field, $B(t) = B_1 \cos(2\pi\ftf t)$, with $B_1 = x_0\frac{\partial B}{\partial x}$, $x_0$ and $\ftf$ being the cantilever oscillation amplitude and frequency, respectively, $\frac{\partial B}{\partial x}$ the field gradient along the oscillation axis, and $B$ the vector component of the magnetic field along the NV anisotropy axis.
	
	In the absence of a magnetic field gradient, the luminescence signal is $R(t)$, as per Eq.~\eqref{eq:Rt_continuous}. With a non-zero field gradient, $R(t)$ incurs an additional phase modulation,
	\begin{equation}
		R'(t) = R\left(t + \frac{\Delta \phi}{2\pi \fmod}\cos\left(2\pi\ftf t \right)\right).
		\label{eq:gradPM}
	\end{equation}
	Here the phase modulation depth is $\Delta\phi = \ye B_1/\dfwin$.
	In the limit of a small gradient ($\Delta\phi\ll 1)$, we can expand Eq.~\eqref{eq:gradPM} to first order,
	\begin{equation}
		R'(t) \approx R(t) + \frac{\mathrm d R(t) }{\mathrm dt} \frac{\Delta \phi}{2\pi \fmod}\cos\left(2\pi\ftf t \right).
		\label{gradPM_smallangle}
	\end{equation}
	The first term is simply the zero-gradient signal, from which we extract the static field as by demodulation at $\fmod$. The second term is an amplitude modulation of $\mathrm d R(t) /\mathrm dt$ at the frequency $\ftf$. We thus observe an up-converted version of $\mathrm d R(t) /\mathrm dt$ centered on $\ftf$.
	
	Next, we express $R(t)$ as a Fourier series,
	\begin{align}
		R(t) = \sum_{k=-\infty}^{\infty} a_k \exp\left(2\pi i k\fmod t\right) \ ,
	\end{align}
	where $a_k\sim \exp(-\pi|k|/\alpha)$, see Eqs.~\eqref{eq:ft_lorentzian} to \eqref{eq:a2_approx}.
	The time derivative is given by
	\begin{align}
		\dot R(t) = 2\pi \fmod \sum_{k=-\infty}^{\infty}  ik a_k \exp\left(2\pi i k\fmod t\right) \ .
	\end{align}
	We next rewrite the second term in Eq.~\eqref{gradPM_smallangle} as
	\begin{align}
		&\dot R(t) \frac{\Delta \phi}{2\pi \fmod}\cos\left(2\pi\ftf t \right) \\
		=& \dot R(t) \frac{\Delta \phi}{2\pi \fmod}\Re\left[\exp\left(2\pi i\ftf t \right)\right] \\
		=& \frac{\Delta \phi}{2\pi \fmod} \cdot \Re\left[\dot R(t) \exp\left(2\pi i\ftf t \right)\right] \\
		=& \Delta \phi \cdot \Re\left[ \sum_{k=-\infty}^{\infty}  ik a_k  \exp\big(2\pi i (\ftf + k\fmod)\big)\right].
	\end{align}
	Therefore, we expect spectral components at $\ftf + k\fmod$.  Since $a_k$ are known (by demodulation at $k\fmod$), measuring the amplitude of the $\ftf + k\fmod$ components will enable the inference of $\Delta\phi$, and thus the gradient strength. Alternatively, the phase of the gradient signal itself also contains useful information about the static field, so the measurements of $k\fmod$ and $\ftf + k\fmod$ harmonics may be combined into a single estimate of both the field and the gradient.

\end{document}